\documentclass[aps,amsmath,twocolumn,amssymb,floatfix,showpacs,
superscriptaddress,footinbib,longbibliography,prb,reprint]{revtex4-1}
\pdfoutput=1
\usepackage{graphics}
\usepackage{bm}
\usepackage{epsfig}
\usepackage{enumerate}
\usepackage{subfigure}
\usepackage{amsmath}
\usepackage{color}
\usepackage{braket}
\usepackage{graphicx}
\usepackage{hyperref}
\hypersetup{
    colorlinks=true,
    linktoc=page,
    linkcolor=red,
    citecolor=blue,
    urlcolor=magenta
}
\usepackage[utf8]{inputenc}

\begin{document}

\title{Generation of concurrence in a generalized central spin model with a three-spin interacting environment}  

\author{Adithya A. Vasista}
\email[Both the authors contributed equally]{}
\affiliation{Department of Physics, BITS Pilani-Hyderabad Campus, Telangana 500078, India}
\author{Anushka Agrawal}
\email[Both the authors contributed equally]{}
\affiliation{Department of Physics, BITS Pilani-Hyderabad Campus, Telangana 500078, India}
\author{Tanay Nag}
\email[Corresponding author: ]{tanay.nag@hyderabad.bits-pilani.ac.in}
\affiliation{Department of Physics, BITS Pilani-Hyderabad Campus, Telangana 500078, India}

\begin{abstract}
We consider the three-spin Ising model to study the effect of three-spin interacting term on bi-partitie entanglement between adjacent spins. The three-dominated disordered region has tri-partite entanglement causing a vanishingly small concurrence, while it acquires maximum value around the critical points. Considering the above model as an environment, we construct a generalized central spin model where two central spins, initially in an unentangled pure state, are coupled locally to two distinct sites of the environmental spin chain. We study the generation of mixed state entanglement between the central spins when the transverse field of the environment is kept fixed,  and suddenly quenched, referring to equilibrium and non-equilibrium dynamics of the central spins, respectively. For the critical environment in the equilibrium, the concurrence shows a dip-revival structure governed by quasi-particle movement. In the non-equilibrium study, we find an initial growth of concurrence followed by a two-stage fall for the inter-phase quench, which is governed by dynamic decoherence channels. The central spins are maximally entangled for a quench in the vicinity of a multicritical point, which arises due to three-spin interaction only. The concurrence becomes long-lived for an intra-phase quench, and this sustainability depends on the strength of the three-spin interaction. Therefore, the three-spin interaction indeed helps in generating bi-partite entanglement in the central spins. 

\end{abstract}
\maketitle
\section{Introduction}
\label{sec1}

The quantum information theoretic measures are found to be very useful to understand and characterize the quantum critical phenomena which is the prime interest of statistical physics \cite{PhysRev.47.777,PhysRevA.40.4277,PhysRevA.54.3824,dutta2015quantum,dutta2010quantum,polkovnikov2011colloquium,RevModPhys.83.863,sachdev1999quantum,vedral2006introduction,chakrabarti1996quantum}. This is due to the fact that the quantum correlations present in the quantum critical systems, in particular, quantum phase transitions,  are caused by the underlying entanglement profile of the quantum states \cite{osterloh2002scaling,PhysRevA.66.032110,PhysRevLett.93.250404,chen2006sublattice,PhysRevLett.93.086402,RevModPhys.80.517,RevModPhys.81.865,PhysRevLett.88.017901,RevModPhys.75.715}. There exist various quantum information theoretic measures e.g., quantum fidelity \cite{PhysRevE.74.031123,PhysRevLett.99.095701,PhysRevLett.99.100603}
entanglement entropy \cite{PhysRevLett.90.227902,latorre2003ground,calabrese2004entanglement,PhysRevE.97.042108}, concurrence \cite{PhysRevLett.78.5022,PhysRevLett.80.2245,wootters2001entanglement,PhysRevLett.93.250404,PhysRevA.80.032304}, and
quantum discord \cite{PhysRevLett.88.017901,PhysRevB.78.224413,PhysRevA.77.042303,PhysRevA.80.022108,nag2011quantum} which are able to capture the
ground state singularity associated with the quantum phase transition. To be precise, bi-partite entanglement is conventionally quantified using separability  (measurement) quantum information-theoretic measures, namely, concurrence (quantum discord). On the other hand,  the entanglement between subsystems of a composite system is investigated using entanglement entropy that is evaluated via the von Neumann entropy of the reduced density matrix corresponding to one of the subsystems. The derivative of concurrence and quantum discord exhibits universal feature around a quantum critical point, while fidelity and entanglement entropy themselves scale with system size.

In the context of quantum information theoretic measures, the decoherence factor or Loschmidt echo is found to be an important marker that is equivalent to 
the quantum fidelity as far as the characterization of a quantum critical point is concerned \cite{PhysRevLett.91.210403,PhysRevE.96.022136}.   
The central spin model (CSM), comprising a single spin-$1/2$  globally coupled to a many-body environmental spin chain, constitutes a fundamental prototype for exploring quantum decoherence where one can study the loss of coherence of the central spin when the environmental experiences a quantum phase transition. 
The CSM has been systematically investigated under both equilibrium \cite{PhysRevA.72.052113,PhysRevA.75.032337,PhysRevLett.96.140604,sharma2012study} and non-equilibrium
\cite{roy2013fidelity,Nag12,Nag16,Suzuki16,mukherjee2019dynamics,PhysRevA.83.062104,nag2013quenching,PhysRevB.86.020301,mukherjee2019dynamics,patel2013quench}, revealing rich dynamical behavior near quantum critical regions with universal scaling relations. 
Motivated by the interplay between criticality, decoherence through CSM, one can  introduce two non-interacting central spins, each locally or globally coupled to a many-body spin chain that acts as the environment \cite{RevModPhys.75.715,PhysRevLett.96.140604,PhysRevA.75.032337,PhysRevA.75.032333,PhysRevA.81.022113,PhysRevLett.107.010403,PhysRevA.83.062104,PhysRevA.76.042118, Nag16b}.
This generalized central spin model (GCSM) allows us to analyze the emergence and temporal evolution of quantum correlations, specifically concurrence and quantum discord between the central spins during the unitary dynamics of the composite system \cite{PhysRevA.78.022327,PhysRevA.74.054102,PhysRevA.75.062312,Nag16b, PhysRevA.74.054102,PhysRevA.82.062119}.

 
The loss of purity and the generation/degradation of entanglement are studied for the CSM and GSCM models, where the environmental spin chain is considered to be an integrable transverse field Ising model (TFIM) or transverse field XY model in most of the studies. While most previous studies of central-spin dynamics rely on integrable environments such as the transverse-field Ising or XY chains, here we consider a three-spin Ising interaction, which breaks free-fermion integrability. This allows us to examine the robustness of entanglement-based dynamical signatures beyond exactly solvable limits, using exact diagonalization. Given the above background, we seek answers to the following questions: What is the role of the three-spin interaction 
in an integrable environmental spin chain?  How does three-spin interaction mediated multi-critical point affect the dynamics of entanglement? To answer the above questions systematically, we first study the concurrence of the isolated three-spin Ising model where the bi-partite entanglement becomes vanishingly small in the three-spin dominated region.  Bi-partite entanglement acquires significant values around the Ising critical lines. We next examine the effect of the three-spin interaction when two central spins, initially residing in a pure unentangled state, are connected locally to the three-spin Ising model, serving as the environment. This leads to a local change in the transverse field that eventually causes multiple decoherence channels leading to a mixed entangled state.  We show that the critical nature of the environment can be captured by a significant dip in the concurrence at a certain time that is governed by  the quasi-particle interference.  Next, the transverse field is globally changed in a sudden manner to study the tunability of the concurrence generation with various quench paths i.e., entanglement growth and fall. We find multi-critical quench leads to the maximum generation of concurrence while intra-phase quench leads to long-lived concurrence. The critical quench also exhibits distinct behavior with three-spin interaction strength that can be understood from the behavior of concurrence in the isolated model. We also highlight the effect of the three-spin on concurrence by reversing the sign for the final value of the transverse field.

The paper is organized as follows: We describe the three-spin interacting Ising model and the formulation of concurrence in Sec.  \ref{sec2}. We also demonstrate the behavior of concurrence in the isolated spin chain as well. In Sec. \ref{sec3}, we extensively discuss the GCSM and generation of concurrence between the central spins under equilibrium and non-equilibrium conditions when the transverse field is unaltered and suddenly altered, respectively. We analyze the results under various quench paths and reversal of the sign in the final value of the transverse field. We also scrutinize different decoherence channels. At the end, in Sec. \ref{sec4}, we conclude with future directions.

\section{Three-spin Ising chain and concurrence}
\label{sec2}

\subsection{Model}
\label{sec2.1}
The Hamiltonian for the three-spin interacting Ising model is given by \cite{divakaran2007effect,Bhattacharjee18,Zhang15,kopp2005criticality}
\begin{equation}
H = -\frac{1}{2} \Bigg\{ \sum_i \sigma_i^z \left[ h + J_3 \sigma_{i-1}^x \sigma_{i+1}^x \right] + J \sum_i \sigma_i^x \sigma_{i+1}^x \Bigg\}
\label{eq:tsim-Ham}
\end{equation} 
where \( h \) is the transverse field, \( J_3 \) is the strength of the three-spin interaction, and \( J \) controls the nearest-neighbor coupling. The model Eq. (\ref{eq:tsim-Ham}) in the absence of $J_3$ reduces to the  standard TFIM  \cite{Sachdev_2011}, and hence $J_3$ introduces additional criticality leading to a variety of magnetic ordered phases.  To analyze the model Hamiltonian Eq. (\ref{eq:tsim-Ham}), we apply the Jordan-Wigner transformation to map the spin operators to fermionic operators. The transformation is defined as
\begin{eqnarray}
c_i^\dagger &=& \sigma_i^+ \exp\left(-i\pi \sum_{j=1}^{i-1} \sigma_j^z \right), ~\sigma_i^x = 2c_i^\dagger c_i -1, \nonumber \\
c^\dagger &=& \frac{\sigma^x + i\sigma^y}{2}, \quad c = \frac{\sigma^x - i\sigma^y}{2}, \nonumber 
\end{eqnarray}
which satisfy the standard fermionic anti-commutation relations: $
\{c_i^\dagger, c_j\} = \delta_{ij}, \quad \{c_i^\dagger, c_j^\dagger\} = \{c_i, c_j\} = 0$. This
allows the Hamiltonian to become 
quadratic in terms of fermionic operators. One can obtain a $2$-level Hamiltonian in the basis $(c_k,c^{\dagger}_{-k})$ after  Fourier transformation, as given by 
 \cite{divakaran2007effect}
\begin{eqnarray}
      H &=& -\sum_{k>0} (h + \cos k - J_3 \cos 2k)(c_k^\dagger c_k + c_{-k}^\dagger c_{-k}) \\
&+& i(\sin k - J_3 \sin 2k)(c_k^\dagger c_{-k}^\dagger + c_{-k} c_k). 
\label{eq:tsim-Ham2} 
\end{eqnarray}
It can be diagonalized using a Bogoliubov transformation 
$
H = -\sum_k \epsilon_k \eta_k^\dagger \eta_k,
$ whereas  $\eta_k$ are the Bogoliubov quasiparticles and $\epsilon_k$ is the excitation energy or gap given by $\epsilon_k = ( h^2 + 1 + J_3^2 + 2h\cos k - 2h J_3 \cos 2k - 2J_3 \cos k )^{1/2}$. We set $J=1$ for all our calculations.

\begin{figure}[ht]
\centering
    \centering
\includegraphics[width=0.45\textwidth]{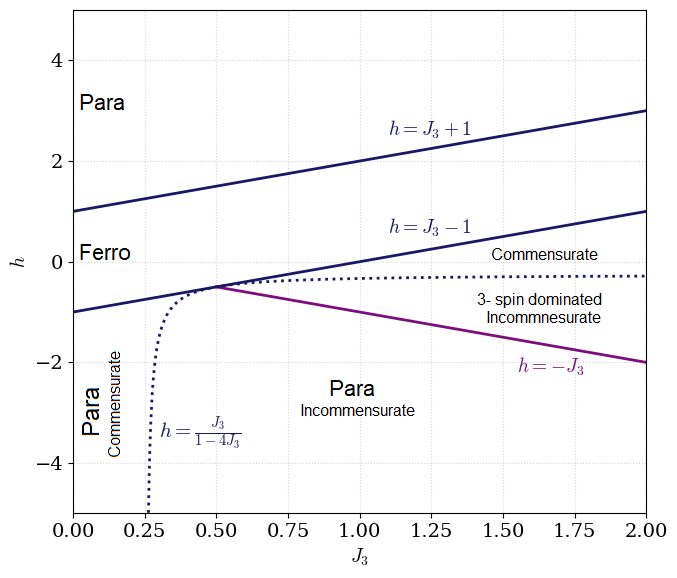}
    \caption{Equilibrium phase diagram of the three-spin interacting Ising model Eq. (\ref{eq:tsim-Ham}). 
    Solid lines show the Ising (three-spin dominated) phase boundaries $h=J_3\pm 1$ ($h=-J_3$), and the dotted line marks the boundary between the incommensurate and commensurate phases.}
    \label{fig:phase}
\end{figure}

We  now show the phase diagram in Fig. \ref{fig:phase} where 
the gap of the spectrum vanishes at $h = J_3 + 1$ and also at $h = J_3 - 1$ with critical wave vectors $k_c=\pi$ and 0, respectively. Across these quantum critical lines,  quantum phase transitions take place between  a ferromagnetically ordered phase and a paramagnetic phase, with exponents the same as those of a TFIM. The critical 
wave vector $k_c$ varies continuously from  $k_c=0$ to $k_c=\pi$ over the critical line $h=-J_3$ with $J_3>0.5$. Note that $h=J_3 \pm 1$ critical line belongs to the  Ising universality class, while $h=-J_3$ corresponds to anisotropic universality class as observed in the transverse XY-model. The paramagnetic phase $h<J_3-1$ comprises commensurate and incommensurate phases within which $3$-spin dominated region is also found to exist. We solve the model using exact diagonalization with $N=20$ spins, where the Hilbert space basis is spanned by $ |s_1\rangle \otimes |s_2\rangle \otimes \cdots \otimes |s_{N-1}\rangle \otimes |s_N\rangle$ where $s=\uparrow,\downarrow$.

\subsection{Concurrence}
Concurrence is a widely used quantum information theoretic measure to quantify entanglement in a two-qubit system. For a pair of adjacent spins located at $i$- and $i+1$-th sites described by a mixed state density matrix \( \rho \), the concurrence \( C(\rho) \)   \cite{wootters2001entanglement,osterloh2002scaling} is computed using the following procedure
\begin{equation}
C(\rho)= {\rm max}(0,\lambda_1 - \lambda_2 - \lambda_3 - \lambda_4)
\label{eq:cnc}
\end{equation}
where $\lambda$'s are eigenvalues of a Hermitian matrix $R=\sqrt { \sqrt { \rho } \tilde{ \rho } \sqrt{\rho} }$ such that $\lambda_1$ $>$   $\lambda_2$ $>$   $\lambda_3$ $>$   $\lambda_4$. Note that $R$ is constructed from the spin-flipped state $\tilde{\rho}= (\sigma_y   \otimes \sigma_y)\rho^*(\sigma_y   \otimes \sigma_y)$.
Since concurrence is defined only for two-qubit systems, we reduce the full many-body density matrix to a $4\times4$ reduced density matrix of two adjacent spins by tracing out the remaining $N-2$ spin degrees of freedom from the $N$ spin density matrix $\rho_N$. This is done using the ground-state wavefunction obtained from the model Hamiltonian Eq. (\ref{eq:tsim-Ham}) after exact diagonalization. \textcolor{black}{ In practice, for numerical stability when dealing with the mixed-state density matrices of the central spin model Sec \ref{sec3}, we compute $\lambda_i = \sqrt{\mu_i}$ where $\mu_i$ are the eigenvalues of the matrix $\rho\tilde{\rho}$~\cite{Nag16}, rather than implementing the nested square roots of Eq.~\eqref{eq:cnc} directly; the two approaches are mathematically equivalent but the former is more robust against numerical noise.}

\subsection{Results}

\begin{figure}[ht]
\centering
\includegraphics[width=0.45\textwidth]{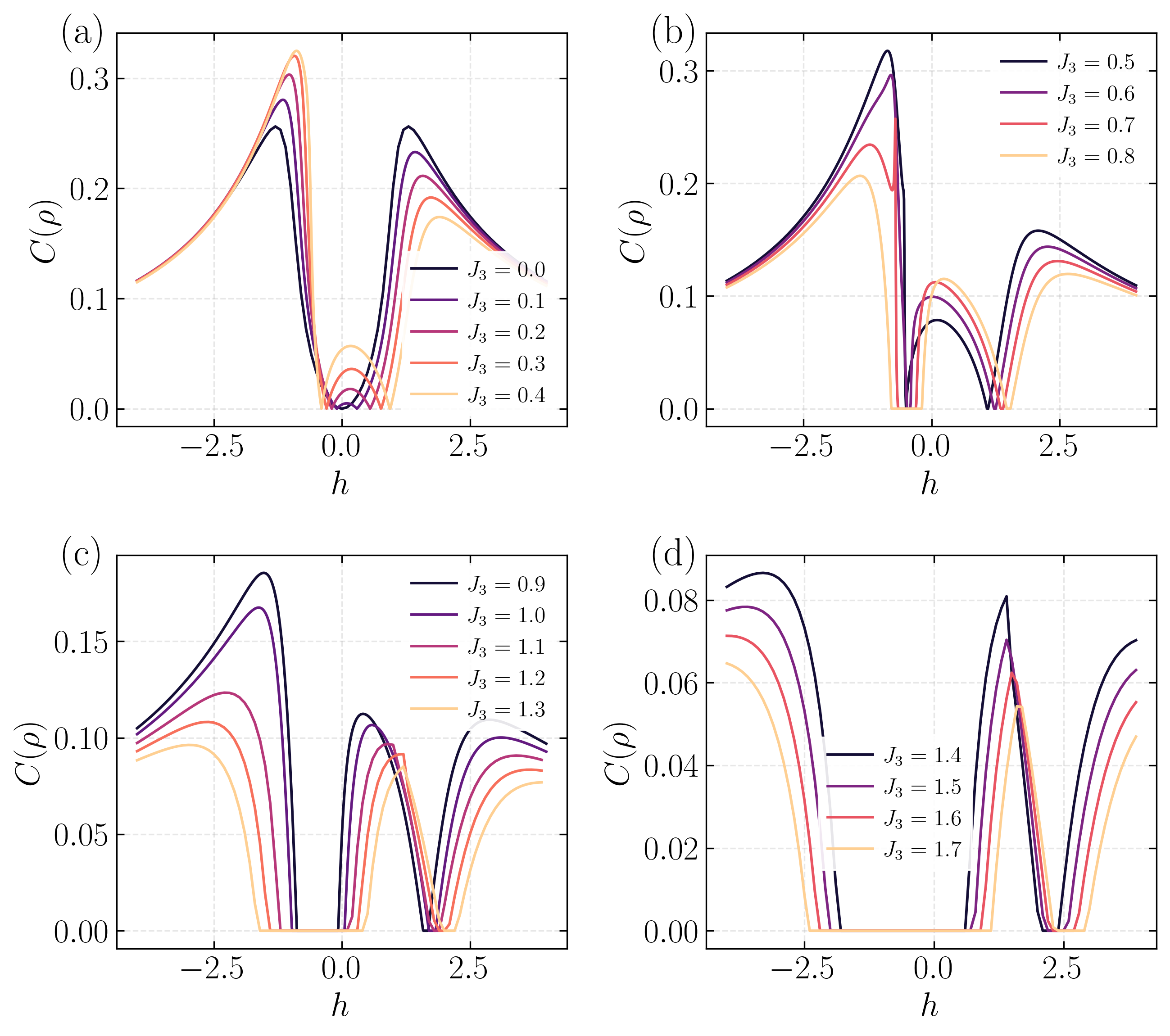}
\caption{Plot for concurrence $C(\rho)$ between nearest neighbour spins, computed using Eq. (\ref{eq:cnc}), as a function of transverse field $h$ in the  three-spin interacting Ising model Eq. (\ref{eq:tsim-Ham})  for $J_3 = 0 $ to $1.7$. We consider $N=20$ for exact diagonalization calculation. }
\label{fig:phase-diagramconc1}
\end{figure}

We analyze quantum phase transitions through the lens of entanglement by examining concurrence as a function of the transverse field \( h \) as shown in Fig.~\ref{fig:phase-diagramconc1}.  This allows clearer identification of phase boundaries, especially in regions with commensurate and incommensurate order.  The concurrence exhibits pronounced peaks at specific values of \( h \), indicating the critical points in the phase diagram.
In Fig.~\ref{fig:phase-diagramconc1} (a), for \( J_3 = 0 \), \textcolor{black}{the concurrence peak corresponds to QCP \( h \approx \pm 1 \), consistent with the critical points of the TFIM.} The entanglement gets maximized at the critical point, resulting in a peak in the concurrence profile \cite{Sachdev_2011}. 
Once the three-spin interaction is introduced through \( J_3 \), the peaks of \textcolor{black}{concurrence shift hence the QCPs shift to $h \approx J_3 +1$ and $h \approx J_3 -1$, and as $J_3$ increases, the left peak tends towards $h=0$ from the negative side and the right peak moves away from $h=0$ to the positive side. We see a small peaks in concurrence which corresponds to QCP $h \approx \frac{J_3}{1 - 4J_3}$ for $J_3$ around $0.25$. As \( J_3 \) increases, when   $J_3 >0.5$, see   Fig.\ref{fig:phase-diagramconc1} (b), the concurrence peak shifts to more negative values of $h$ following \( h = -J_3 \) critical line of the phase diagram. There exist intermediate peaks around $h=0$ bearing the signature of the phase boundary between incommensurate and commensurate phases. However, the critical line $h = J_3-1$ also impacts the concurrence in this intermediate window.  The right side peaks are associated with $h = J_3+1$ phase boundary.} 
 
Importantly, the three-spin dominated incommensurate phase yields vanishingly small concurrence. 
Upon increasing $J_3$ further, as shown in Figs. \ref{fig:phase-diagramconc1} (c,d), we find the region with zero concurrence increases, representing the widened nature of the three-spin dominated incommensurate phase. The left (right) most peak is caused by $h= -J_3$ ($h = J_3+1$) critical line, while the intermediate peak is dominated by $h = J_3-1$ critical line as $J_3$ increases.

It is important to note that the concurrence peaks at $h$ values that are near the critical points \cite{osterloh2002scaling,PhysRevA.66.032110}. This deviation is caused by the finite size of the three-spin model. What is really interesting here is that the entanglement between two qubits gradually decreases as the three-spin interaction term $J_3$ increases. This feature is observed when the value of concurrence systematically decreases from Fig. \ref{fig:phase-diagramconc1} (a) to Fig. \ref{fig:phase-diagramconc1} (d) for the identical window of $h$. This is 
more clearly observed inside the three-spin interaction dominated region $-J_3<h<J_3-1$ within which bi-partite entanglement between two spins vanishes. One can naively think that spin-chain favors GHZ-like state $|{\rm GHZ}\rangle=1/\sqrt{2}(|\uparrow_1 \uparrow_2 \cdots \uparrow_N\rangle + |\downarrow_1 \downarrow_2 \cdots \downarrow_N\rangle )$ inside the three-spin interaction-dominated region, resulting in a two-spin separable state $\rho_{12}$ after tracing out $N-2$ spins, $\rho_{12}=1/2 |\uparrow_1 \rangle \langle \uparrow_1| \otimes  |\uparrow_2 \rangle \langle \uparrow_2| + 1/2 |\downarrow_1\rangle \langle \downarrow_1| \otimes |\downarrow_2 \rangle \langle \downarrow_2|$. Importantly, concurrence acquires maximum value around the critical line $h=J_3-1$ when $J_3<0.25$ can be connected to the conventional Ising paramagnetic state, while the three-spin interaction-dominated paramagnetic state leads to a minimal amount of bi-partite entanglement.    This reinforces the idea that bi-partite entanglement measures, such as concurrence, can serve as robust probes of quantum criticality even for a three-spin Ising chain. Interestingly, the concurrence is more in the paramagnetic phase as compared to the ferromagnetic phase due to the disordered (ordered) nature of spin moments in the paramagnetic (ferromagnetic) phase.


\section{Concurrence using generalized central-spin model}
\label{sec3}

\begin{figure}[ht]
\centering
\includegraphics[width=0.23\textwidth]{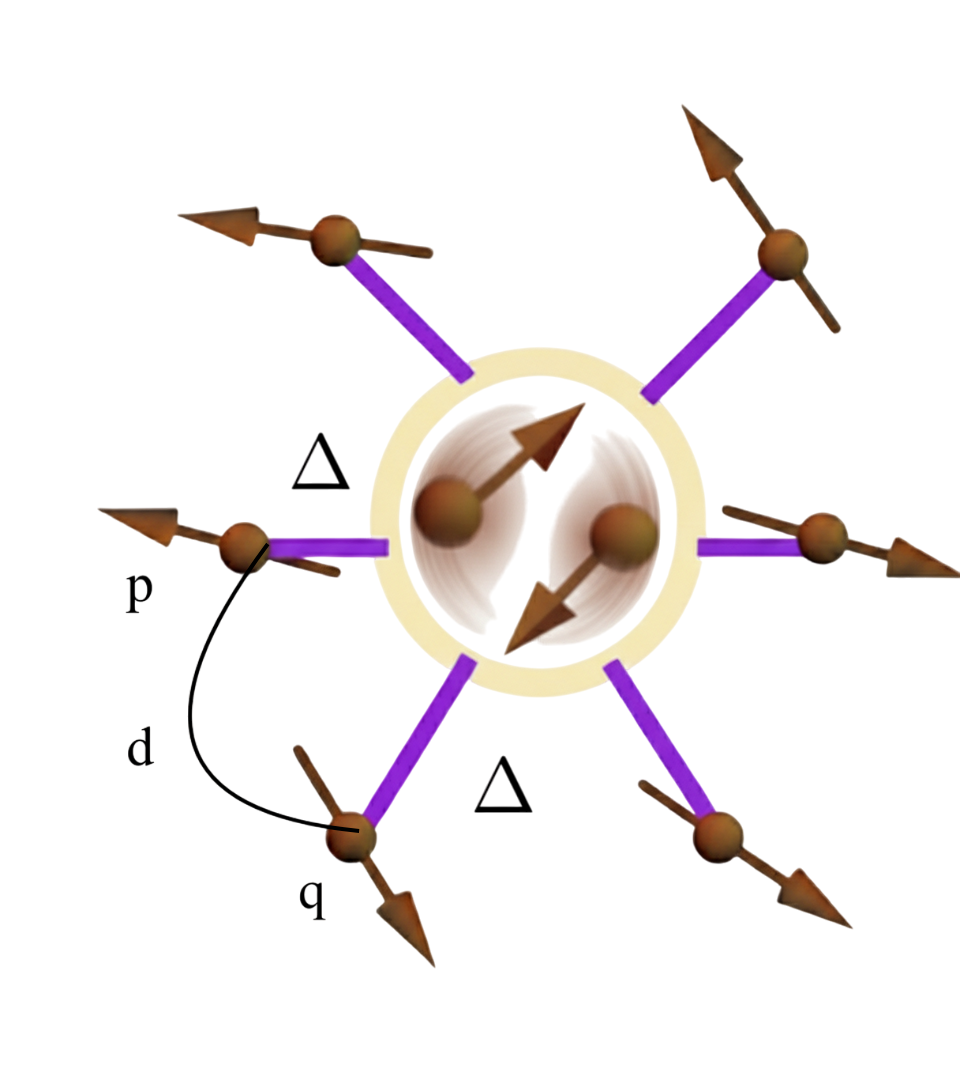}
\caption{Illustration explaining the schematic  GCSM where the central yellow part represents the two-qubit spin system with spins $A$ and $B$. Spin $A (B)$ is coupled locally to site $p (q)$, identified by blue bonds, of the environmental three-spin Ising model.  The central as well as the environmental spins are represented by brown arrows.}
\label{fig:cartoon}
\end{figure}

We consider GCSM where two central spins are locally coupled to the three-spin Ising model serving as the environmental spin chain \cite{Quan_2006}.   The schematic diagram of the GCSM is shown in Fig. \ref{fig:cartoon}. The composite Hamiltonian $H_T$, comprising an environmental three-spin Ising Hamiltonian $H_E$ with $N$ sites given in Eq. (\ref{eq:tsim-Ham}) and interaction Hamiltonian $H_{SE}$ of two spins, is given by $H_T =H_{SE}+H_E$ where we consider  PBC, $\sigma^{i}_{N+1} = \sigma^{i}_1$ for the environmental spin chain as described in Eq. (\ref{eq:tsim-Ham}). The interaction Hamiltonian takes the form $H_{SE}=-\Delta(\ket\uparrow\bra\uparrow_A \otimes \sigma^z_p+\ket\uparrow\bra\uparrow_B \otimes \sigma^z_q)$    where \( A \) and \( B \) denote the two central spins, and \( p \) and \( q \) are their respective coupling sites to the environmental chain, separated by a distance \( d = | p-q | \). The parameter \( \Delta \) controls the interaction strength. $\ket\uparrow_{A,B}$ is an eigenstate of the central spin $\sigma^z_{A,B}$ satisfying the relation $\sigma^z_{A,B} \ket\uparrow_{A,B}= \ket\uparrow_{A,B}$. We note that there is no cross term in $H_{SE}$ allowing for the flip of the central spin. Note that $[H_S,H_{SE}]=0$ further emphasizing the absence of dynamics in the central spins themselves.  
We typically consider the weak-coupling limit \( \Delta \to 0 \), where the quantum phase transitions, taking place in the environmental spin chain, determine the dynamics of the qubits substantially. To study entanglement generation, we initialize the system in a completely unentangled product state of the two central spins $  \ket\phi_{AB} = \frac{1}{2} (\ket\uparrow _A + \ket\downarrow_A)\otimes (\ket\uparrow_B + \ket\downarrow_B)  $. The state of the composite system at initial time $t=0$ is given by $|\Psi (t=0)\rangle_{T}=\ket\phi_{AB}\otimes |\eta(h_I,t=0)\rangle_E$ where $|\eta(h_I,t=0)\rangle_E$ represents the initial ground state of the environmental spin chain with transverse field $h_I$.


One finds that the  composite state evolves as 
$\rho_{T}(t=0)=|\Psi (t=0)\rangle \langle \Psi (t=0)|_T$ under the composite unitary operator $U_{T}= e^{-iH_T t}$ such that $\rho_{T}(t)= U_{T} \rho_{T}(t=0) U^{\dagger}_{T}$. Here $U_T$ includes $H_{SE}$ leading to the environmental evolution conditioned by the state of the central spins. One can think of this evolution as an effective two-qubit CNOT operation where the central spin system acts as a control and the environment plays the role of a target. The state of the target changes depending on the state of the control, namely the central spin system.  To be precise, $(H_{SE} \otimes H_E) (|\alpha_A \beta_B\rangle \otimes |\eta\rangle)=|\alpha_A \beta_B\rangle \otimes H^{\alpha \beta}_E |\eta\rangle$ with $\alpha,\beta =\uparrow, \downarrow$.
We examine two distinct situations: (a) the transverse field of the environment is kept fixed with time i.e., $h_I=h_F=h$, leading to $| \eta (h_I,t) \rangle= U^I_E | \eta (h_I,t=0) \rangle$ representing the equilibrium case  and (b) the transverse field of the environment is suddenly changed from $h_I$ to $h_F$ when time evolves from $t=0$ to $t>0$ leading to $| \eta (h_F,t) \rangle= U^{F}_E | \eta (h_I,t=0) \rangle$
representing the non-equilibrium case. A careful analysis suggests that $U^I_E=\exp(-i H^{\alpha \beta}_I t)$ and $U^{F}_E=\exp(-i H^{\alpha \beta}_F t)$ with 
\begin{eqnarray}
    H^{\alpha \beta}_{I/F}&=& -\frac{1}{2} \Bigg\{ \sum_i \sigma_i^z \big[ h_{I/F} + 2 \Delta \delta_{p,i} \delta_{\uparrow,\alpha} +  2 \Delta \delta_{q,i} \delta_{\uparrow,\beta} \nonumber \\
    &+& J_3 \sigma_{i-1}^x \sigma_{i+1}^x \big] + J \sum_i \sigma_i^x \sigma_{i+1}^x \Bigg\} 
\label{eq:tsim-Ham3}
\end{eqnarray}
More precisely, the environmental time-evolved state is given by 
\begin{equation}
          \ket{\eta_{\alpha\beta}(h_{I/F},t>0)}= e^{-iH^{\alpha\beta}_{I/F}t} \ket{\eta (h_I ,t=0)}
\end{equation}
with $H^{\downarrow\downarrow}_{I/F}= H_E(h_{I/F})$, 
$H^{\uparrow\uparrow}_{I/F}= H_E(h_{I/F}) -\Delta(\sigma_p^z + \sigma_q^z)$,
$H^{\uparrow\downarrow}_{I/F}= H_E(h_{I/F}) -\Delta \sigma_p^z$,
$H^{\downarrow\uparrow}_{I/F}= H_E(h_{I/F}) -\Delta \sigma_q^z$. The time-evolved composite state is given by $\rho^{I/F}_T(t>0)=\frac{1}{4}\sum_{\alpha, \beta,\lambda, \gamma}|\alpha \beta \rangle \langle \lambda \gamma | \otimes \ket{\eta_{\alpha\beta}(h_{I/F},t>0)} \bra{\eta_{\lambda \gamma}(h_{I/F},t>0)} $ with $\alpha, \beta,\lambda, \gamma=\uparrow,\downarrow$. The reduced density matrix of the central spins are given by $\rho_S^{I/F}(t)={\rm Tr}_E[\rho^{I/F}_T(t)]=\frac{1}{4}\sum_{\alpha, \beta,\lambda, \gamma}|\alpha \beta \rangle \langle \lambda \gamma | d^{I/F}_{\alpha \beta,\lambda \gamma}$ with $d^{I/F}_{\alpha \beta,\lambda \gamma}=\bra{\eta_{\alpha\beta}(h_{I/F},t>0)}\eta_{\lambda \gamma}(h_{I/F},t>0)\rangle $.  The density matrix in the basis $\ket{\uparrow \uparrow}, \ket{\uparrow \downarrow}, \ket{\downarrow \uparrow}, \ket{\downarrow\downarrow}$ is given by
\begin{align}
    \rho^{I/F}_S(t) = \frac{1}{4}
\begin{bmatrix}
1 & {d^{I/F}_{\uparrow\uparrow,\uparrow\downarrow}} & {d^{I/F}_{\uparrow\uparrow,\downarrow\uparrow}} & {d^{I/F}_{\uparrow\uparrow,\downarrow\downarrow}} \\
({d^{I/F}_{\uparrow\uparrow,\uparrow\downarrow}})^* & 1 & {d^{I/F}_{\uparrow\downarrow,\downarrow\uparrow}} & {d^{I/F}_{\uparrow\downarrow,\downarrow\downarrow}} \\
({d^{I/F}_{\uparrow\uparrow,\downarrow\uparrow}} )^*& ({d^{I/F}_{\uparrow\downarrow,\downarrow\uparrow}})^* & 1 & {d^{I/F}_{\downarrow\uparrow,\downarrow\downarrow}} \\
({d^{I/F}_{\uparrow\uparrow,\downarrow\downarrow}} )^*& ({d^{I/F}_{\uparrow\downarrow,\downarrow\downarrow}})^* & ({d^{I/F}_{\downarrow\uparrow,\downarrow\downarrow}} )^*& 1
\end{bmatrix}
\label{eq:dm_gcsm}
\end{align}
Note that $\rho^{I}_S(t)$ corresponds to the case (a) of the equilibrium scenario. On the other hand, $\rho^{F}_S(t)$ denotes the case (b) of the non-equilibrium scenario.  Using the above density matrix, one can compute concurrence, quantifying bi-partite entanglement between the two central spins,  as give in Eq. (\ref{eq:cnc}).

It is important to note that the dynamics of the central spin system depend on four environmental evolution channels $d_{\uparrow \uparrow}$, $d_{\uparrow \downarrow}$, $d_{\downarrow \uparrow}$ and $d_{\downarrow \downarrow}$.  We call them as decoherence channels as these channels cause the  purity of the central two-qubit state to reduce, hence, they introduce dephasing to the initial pure density matrix $\rho_S(0)$ leading to a mixed density matrix $\rho^{I/F}_S(t>0)$ at a later time. 
One can directly obtain the 
quantity $ d^{I/F}_{\alpha\beta,\lambda\gamma} $ from the spin-basis of the environmental Hamiltonian given by $ |s_1\rangle \otimes |s_2\rangle \otimes \cdots \otimes |s_{N-1}\rangle \otimes |s_N\rangle$ where $s=\uparrow,\downarrow$.  Under the 
periodic boundary condition of the environmental spin chain, the lattice index $p$, $q$  are reversible among themselves leading to 
$d^F_{\uparrow \uparrow, \downarrow \downarrow}$, $d^F_{\uparrow \uparrow, \downarrow \uparrow}$, $d^F_{\downarrow \downarrow, \downarrow \uparrow}$, and $d^F_{\uparrow \downarrow, \downarrow \uparrow}$ as the four independent channels as  $d^F_{\uparrow \uparrow, \downarrow \uparrow}=d^F_{\uparrow \uparrow, \uparrow \downarrow}$ and $d^F_{\downarrow \downarrow, \downarrow \uparrow}=d^F_{\downarrow \downarrow, \uparrow \downarrow}$.
We already benchmark our results, obtained from exact diagonalization \cite {PhysRevB.42.6561}, with those of using the Jordan-Wigner transformation \cite{ovrum2007quantumcomputationalgorithmmanybody, PhysRevA.94.022316}. We have fixed $\Delta=0.1$ for our upcoming analyses.


\subsection{ Case (a): Equilibrium Study with $h_I=h_F$}
\label{sec3.1}
\begin{figure}[ht]
\centering
\includegraphics[width=0.42\textwidth]{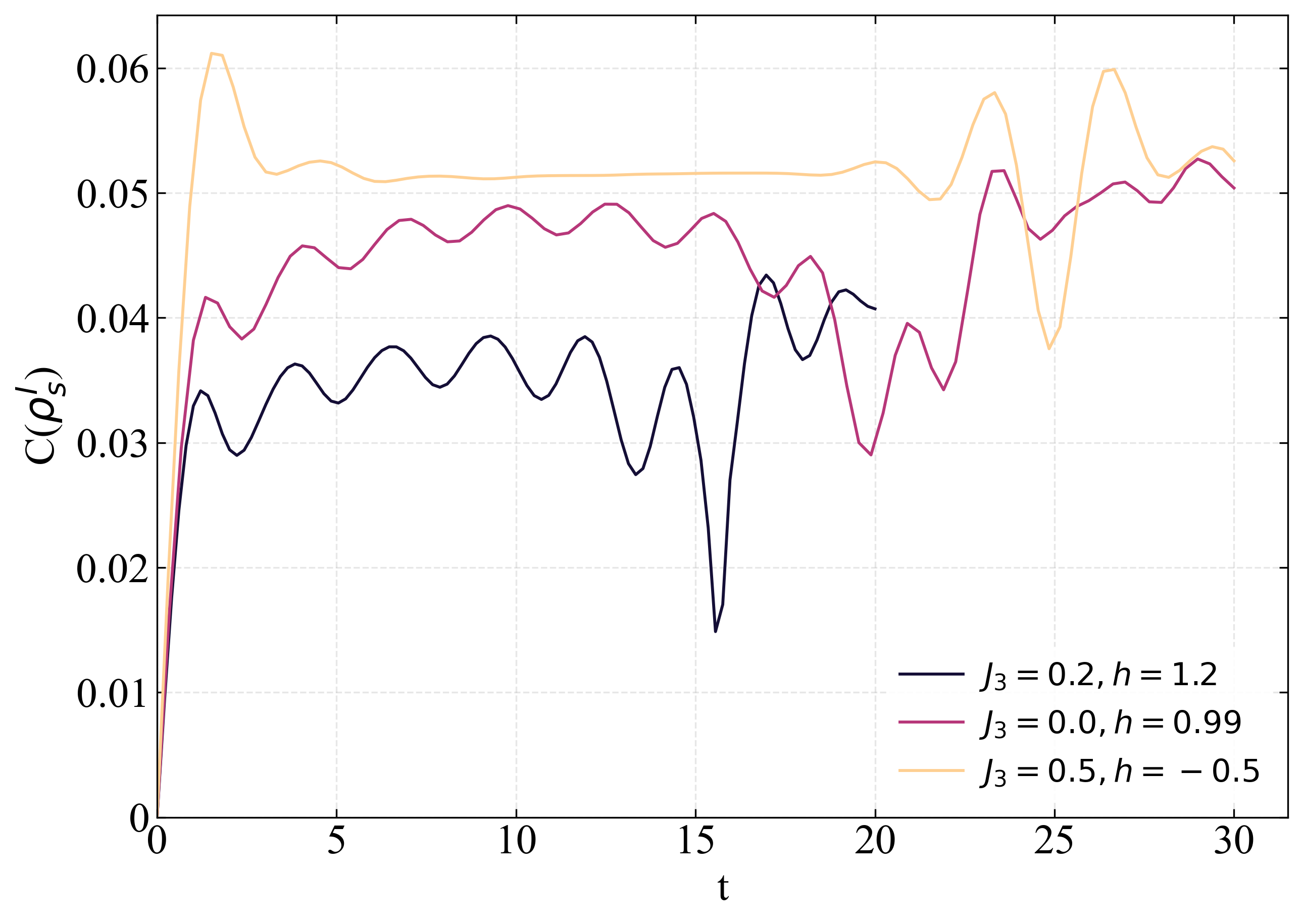}
\caption{Equilibrium plot shows the behavior of system's concurrence $C(\rho_S^I)$ with time on the $h=J_3+1$ critical line for ($J_3=0, h=1.0$, $d=0$ and $J_3=0.2, h=1.2$, $d=0$) and at the multi-critical point ($J_3=0.5, h=-0.5$, $d=0$). We consider $N=20$ spins for exact diagonalization calculation.}
\label{fig:dip_N}
\end{figure}

\begin{figure}[ht]
\centering
\includegraphics[width=0.42\textwidth]{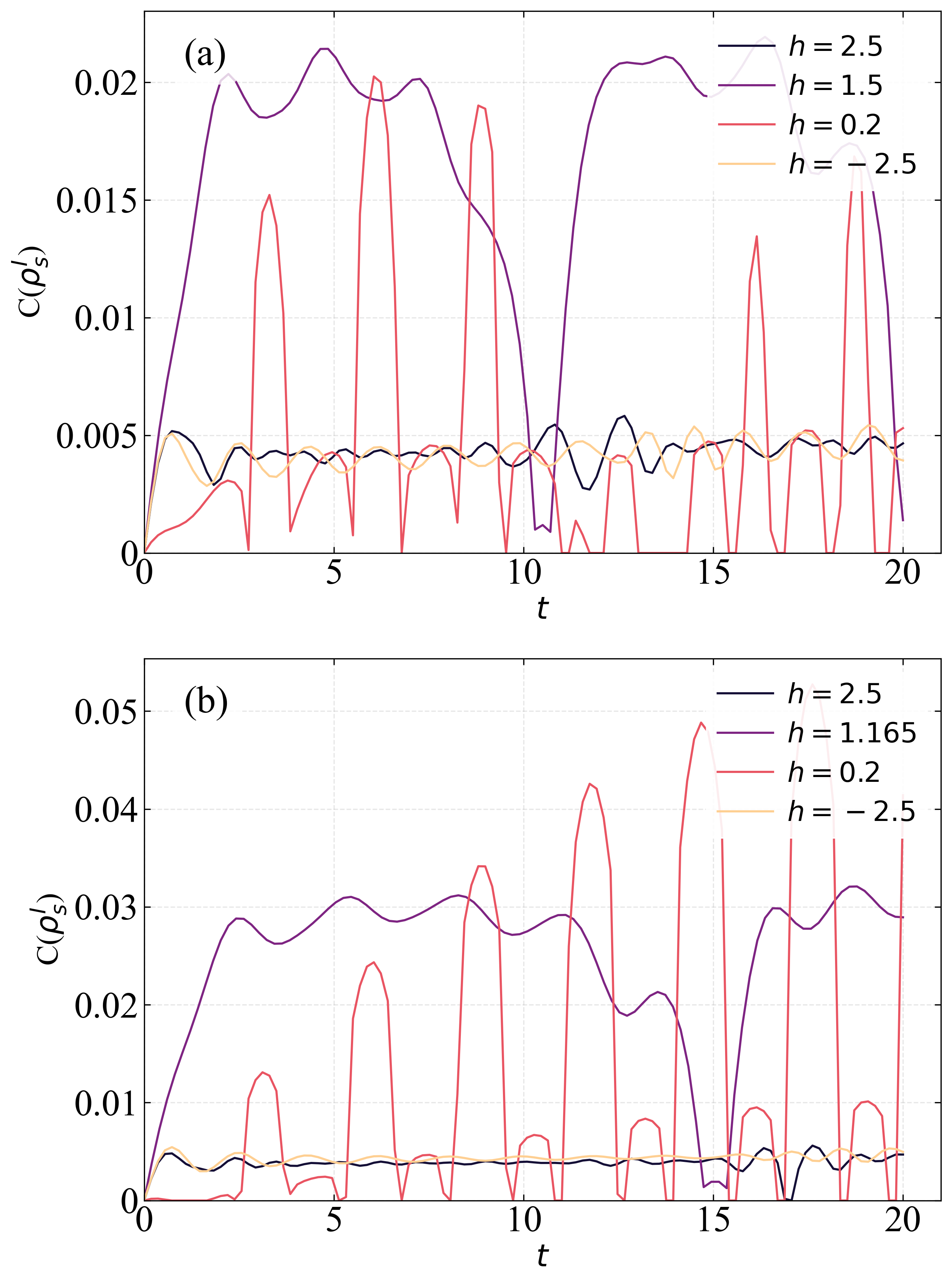}
\caption{
Time evolution of concurrence \( C(\rho_S^I) \) as a function of time at critical points corresponding to \( h = 1 + J_3 \).     
(a) For \( J_3 = 0.5 \), \( h_I = h_F = 1.5 \), a sharp dip is observed at \( t = N/2 \), consistent with coherent quasiparticle interference and linear dispersion with group velocity \( v_{\text{max}} = 2 \).     
(b) For \( J_3 = 0.165 \), \( h_I = h_F = 1.165 \), the concurrence dip occurs near \( t \approx 15 \), corresponding to \( v_{\text{max}} \approx 1.33 \). We consider $d=1$ and $N=20$ spins for exact diagonalization calculation.  } 
\label{fig:equilibriumpredic_collage}
\end{figure}

We now examine the equilibrium dynamics of concurrence $C(\rho^I_S)$, which corresponds to case (a). We now study the concurrence Eq. (\ref{eq:cnc}) computed from the reduced density matrix $\rho^{I}_S(t)$ while the initial and final value of the transverse field are the same $h_I=h_F$.  We show the generation of concurrence as a function of time in Fig. \ref{fig:dip_N} with $h_I=h_F=0.99$ and $J_3=0$.  The concurrence exhibits non-trivial, revival-like dynamics near the critical point, consistent with expectations from a fermionized Hamiltonian with larger system size \cite{PhysRevA.94.022316}. The overall behavior, both qualitatively and quantitatively, resembles that of the fermionized systems. For example, the concurrence oscillations and dip at $t=t_d=N$ are clearly observed following the exact diagonalization method, which are the same as the fermionized results as observed for TFIM \cite{PhysRevA.94.022316}.  The dip at $t=N$ is a trademark signature of critical behavior of concurrence in TFIM, when $J_3=0$. In the presence of three-spin interaction $J_3 \ne 0$, the dip is observed for $h \approx J_3 +1 $ in Fig. \ref{fig:equilibriumpredic_collage}, identifying a deep connection with the group velocity $v_g$ of the quasi-particle. To be precise, $t_d=N/v_g$ where quasi-particles can travel through the periodic environmental spin chain.  We note that the dip height increases with $N$ in a power law fashion.

We know that the group velocity of the quasi-particle is the maximum value $v=\partial \epsilon_k/\partial k$. For $J_3=0$, the group velocity $v_{\rm max}\approx 1$, while for $J_3 \ne 0$, $v_{\rm max}$ depends on $J_3$ as shown in Figs. \ref{fig:equilibriumpredic_collage}. The connection with the central spin leads to a local sudden quench at sites $p$ and $q$. This causes the quasi-particles of opposite momentum to fly away from the local quench center.   These quasi-particles initially move away from each other and finally approach each other due to the periodic nature of the chain. Therefore, the relative distance between the two counterparts of a given pair varies periodically with time, leading to the peak-dip revival structure of the concurrence. Once they are maximally (minimally) separated in the periodic chain, the concurrence acquires peak (dip) value with time. Therefore, the primary peak-dip structure is caused by the global interference of the quasi-particles. 
The smaller oscillations between two successive dips are the artifacts of local interference. The above global interference is a signature of the fact that the correlation length is larger than the open length of the periodic chain. In the off-critical cases, the global interference is absent as the correlation length is smaller than the chain's length, triggering the local interference only. This generates oscillations in concurrence without any prominent peak-dip structure, as shown in Figs. \ref{fig:equilibriumpredic_collage} (a.b).  Interestingly, the dip time, $t_{d}$, aligns with a ballistic light-cone-like spreading of entanglement, reminiscent of the Lieb-Robinson bound \cite{Lieb:1972wy}. Therefore, the bi-partite entanglement between the qubits of the GCSM, quantified by concurrence $C(\rho_S^I)$, is tuned by the critical nature of the environmental spin chain.

For the off-critical nature of the environmental spin chain, the entanglement between two central spins is usually small with time in the paramagnetic phase of the environment as compared to the ferromagnetic phase.  One can find exactly the opposite behavior of concurrence for an uncoupled three-spin interacting spin chain.  Therefore, the disordered nature of the environmental spin chain is not able to induce entanglement between the central spin. In contrast, the ordered nature of the environment is more helpful to mediate bi-partite entanglement. In the Ferromagnetic phase (\( h = 0.2 \)), concurrence exhibits oscillatory behavior with a larger amplitude, indicating the existence of strong decoherence channels $d^I_{\alpha \beta, \lambda \gamma}$ while the ground state of the environment is already ordered.  The decoherence channels for the disordered ground state of the environment are weak comparatively, leading to feeble generation of bi-partite entanglement.  This is analogous to the fact that the fidelity between two ordered states is intended to be more than that of two disordered states.


\subsubsection{Analysis Through Decoherence Channels}
\label{sec3.6}

\begin{figure}[t]
\centering
\includegraphics[width=0.45\textwidth]{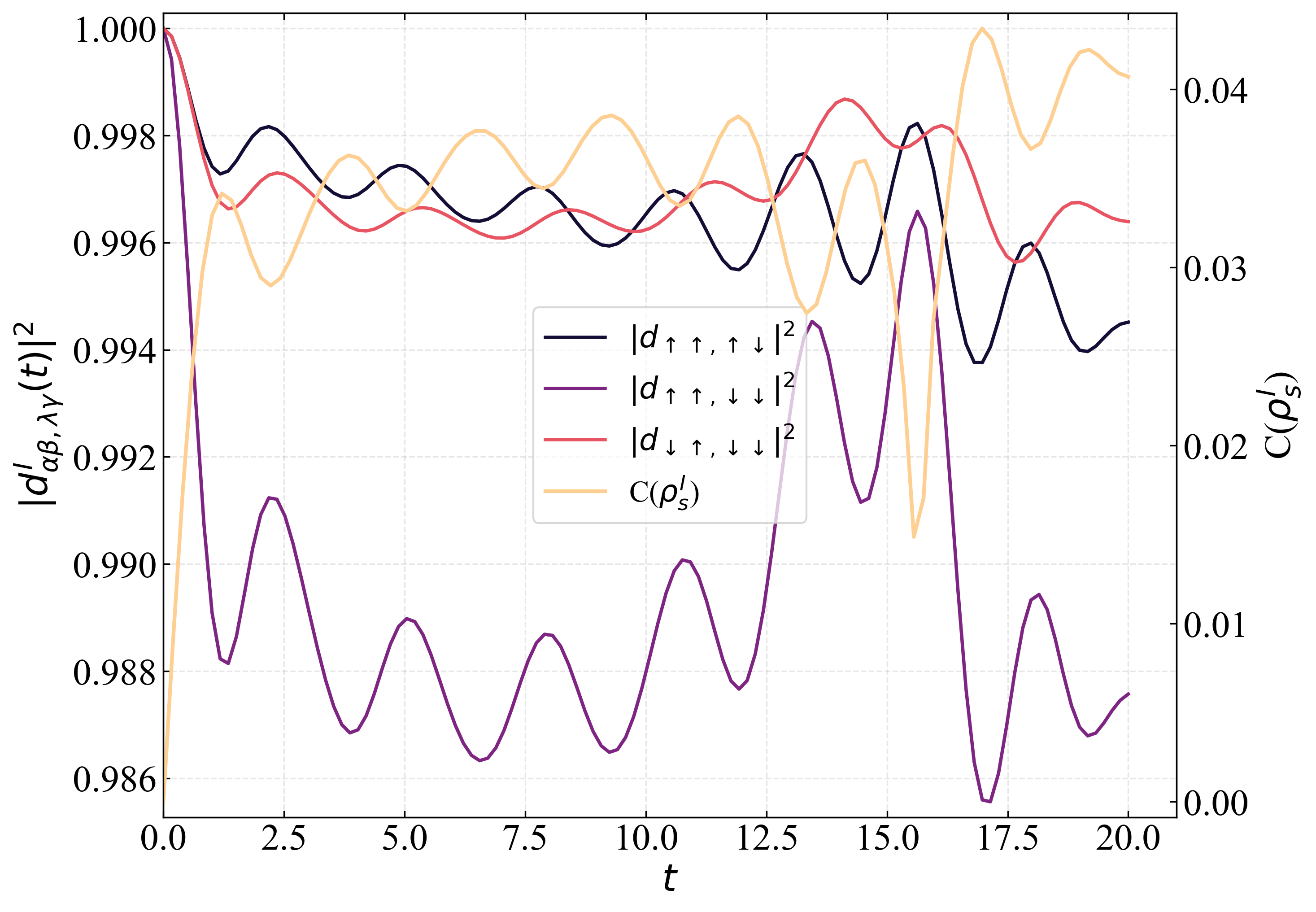}
\caption{Temporal analysis of decoherence channel $d^I_{\alpha \beta, \lambda \gamma}(t)$ (left axis), as shown in Eq. (\ref{eq:dm_gcsm}), and concurrence $C(\rho_S^I)$ (right axis) with time for equilibrium study for $h_I=h_F=1.19$ and $J_3=0.2$.  We consider $d=0$ and $N=20$ spins for exact diagonalization calculation. }
\label{fig:deco}
\end{figure}

We show the evolution of the off-diagonal decoherence channels $|d_{\alpha\beta, \lambda \gamma}|^2$ (left axis) and concurrence $C(\rho_S^I)$ (right axis) in Fig.~\ref{fig:deco} for $h_I = h_F$ with $J_3 = 0.2$ and $d=0$. The fall of these channels is insignificant, leading to a tiny loss of purity and a little entanglement induction. Since $d=0$, the $|\uparrow\downarrow\rangle$ and $|\downarrow\uparrow\rangle$ states couple to the same environmental site $p$, resulting in identical Hamiltonians $H^{\uparrow\downarrow} = H^{\downarrow\uparrow} = H_E - \Delta\sigma_p^z$. Consequently, their decoherence channels $|d_{\uparrow\downarrow, \downarrow\downarrow}|^2$ and $|d_{\downarrow\uparrow, \downarrow\downarrow}|^2$ (black and pink lines) are identical and follow each other. In contrast, the $|\uparrow\uparrow\rangle$ state corresponds to a perturbation of $H^{\uparrow\uparrow} = H_E - 2\Delta\sigma_p^z$. This stronger $2\Delta$ perturbation causes a faster loss of overlap, which is why the $|d_{\uparrow\uparrow, \downarrow\downarrow}|^2$ channel (violet line) drops more significantly than the other two. Interestingly, 
the revival in the channel $|d_{\uparrow\uparrow, \downarrow\downarrow}|^2$ is directly related to the dip in concurrence $C(\rho_S^I)$ that appears around $t \approx 15$.

\subsection{Case (b): Non-equilibrium study with $h_I \ne h_F$}
\label{sec3.3}
We extend the equilibrium study to the non-equilibrium scenario where $h_I$ is suddenly changed to $h_F$ for $t>0$. \textcolor{black}{To investigate the non-equilibrium dynamics, we implement a sudden  quench of the transverse field which is instantaneously modified for all the environmental spins at $t=0+$. 
To be precise, the time-dependent field is described by the Heaviside step function $\Theta(t)$ as
\begin{equation*}
    h(t) = h_{I} + (h_{F} - h_{I})\Theta(t) = 
    \begin{cases}
        h_{I}, & t \leq 0, \\
        h_{F}, & t > 0
    \end{cases}
\end{equation*}
where $\Theta(t)=1$ for $t>0$ elsewhere $\Theta(t)=0$.  
At $t \leq 0$, the environmental spin chain is prepared in the ground state $|\eta(h_{I})\rangle_{E}$ of 
the Hamiltonian $H_{E}(h_{I})$. At $t = 0$, the field is instantaneously switched to $h_{F}$, and for 
$t > 0$ the system evolves unitarily under the new Hamiltonian $H_{E}(h_{F})$ according to 
$|\eta(h_{F},t)\rangle = e^{-iH_{E}(h_{F})t}|\eta(h_{I})\rangle_{E}$. This discontinuous change creates 
a non-equilibrium scenario where the initial state is no longer an eigenstate of the final Hamiltonian, 
leading to coherent dynamics and distinct entanglement generation compared to the equilibrium case.}

\begin{figure}[ht]
\centering
          \includegraphics[width=0.45\textwidth,height=13.5cm]{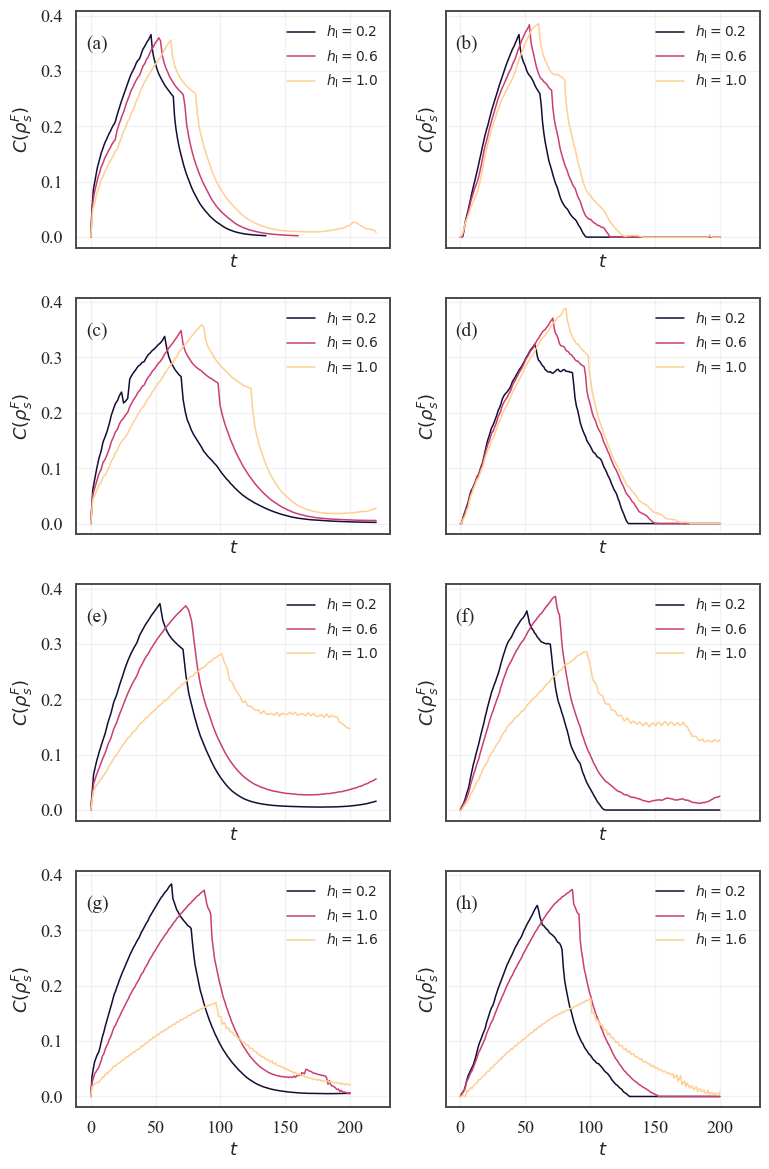}
\caption{We show the dynamics of concurrence $C(\rho_S^F)$, computed from $\rho_S^F(t)$ Eq. (\ref{eq:dm_gcsm}) for different three-spin interaction strengths \( J_3=0.1 \), $0.6$,$1.5$,$2.0$ in (a,b), (c,d), (e,f), (g,h) respectively under a quench from $h_I=0.2$, $0.6$, $1$,$1.6$ to \( h_F = 2.2 \) to highlight different regimes of inter- and intra-phase queching schemes.    We show $d=0$ and $5$ behavior in (a,b,c,d) and (e,f,g,h), respectively. We consider $N=20$ spins for exact diagonalization calculation. }  
\label{fig:Dynamic_collage}
\end{figure}

This is a global quench for the environmental spin chain in addition to the local quench, caused by the GCSM, which is seen in the equilibrium case. We start with Figs.  \ref{fig:Dynamic_collage} (a,b,c,d,e,f,g,h) where we systematically study the effect of $J_3$ and the distance $d$ between the coupled spins, provided that the final value of the transverse field is kept fixed at $h_F=2.2$. We find the following generic features for inter-phase quench crossing the critical point. The concurrence increases as a function of time in a sub-diffusive way with $C\propto t^{\alpha}$, where our numerical fits find the growth exponent $\alpha$ is typically in the range $0.5<\alpha<0.76$, depending on the quench parameters. The concurrence inter-phase quenching shows a two-stage super-diffusive fall with time as $C\propto t^{-\beta_{1}}$ in the primary fall, followed by a secondary fall $C\propto t^{-\beta_{2}}$. Our fits confirm this two-stage behavior, yielding a primary fall exponent $\beta_1$ in the range of $1.5<\beta_1<2.5$ and a much faster secondary fall exponent $\beta_2$ in the range of $3<\beta_2<7$. This two-stage drop causes the concurrence to display a primary peak and a secondary peak at times $t=t_1$ and $t=t_2$, respectively. The time at which concurrence exhibits these peaks increases as $h_I$ increases. The central qubits turn into a mixed entangled state similar to the Werner state as $t>0$. Followed by the secondary fall, the two-qubit state no longer remains substantially entangled. The rise and fall of concurrence are caused by constructive and destructive interference between different decoherence channels, which we discuss below in detail. We investigate the growth and decay profile of intra-phase quenching extensively.

Interestingly,  as soon as the two-qubit state becomes completely mixed $\rho^F_S(t\gg t_2)=I/4$, the concurrence vanishes, leading to a separable state $\rho^F_S(t\gg t_2)=\rho_A \otimes \rho_B$. One can assume the long-time mixed  unentangled state as $\rho^F_S(t)=p_{\uparrow \uparrow}\rho_{\uparrow\uparrow}+ p_{\uparrow \downarrow}\rho_{\uparrow \downarrow} + p_{\downarrow \uparrow}\rho_{\downarrow \uparrow}+ p_{\downarrow \downarrow}\rho_{\downarrow \downarrow}$ with $p_{\downarrow \downarrow} =p_{\downarrow \uparrow}=p_{\uparrow \downarrow}=p_{\uparrow \uparrow}=1/4$, $\rho_{x y}=|{xy}\rangle \langle {xy}|$, $x,y \in \{\uparrow,\downarrow\}$. In this extreme time limit, 
all the off-diagonal terms in $\rho^F_S$ vanish. Now, the primary and secondary peaks in concurrence are connected to the vanishing nature of the different decoherence channels \cite{PhysRevA.94.022316}.  The different channels behave independently, while they vanish at distinct times, leading to the two-stage fall of the concurrence. It has been shown for environmental TFIM  that $d^F_{\uparrow \uparrow, \downarrow \downarrow} \to 0$ mediates the primary peak to appear while $d^F_{\uparrow \uparrow, \downarrow \uparrow} d^F_{\downarrow \downarrow, \downarrow \uparrow}  \to 0$ causes the secondary peak to appear \cite{PhysRevA.94.022316}. \textcolor{black}{The same is expected to qualitatively hold for the present case of the three-spin Ising model environment as well. Importantly, similar to the TFIM environment,  $|d^F_{\uparrow \uparrow, \downarrow \downarrow}(t)|^2$ remains the most rapidly decaying channel in the presence of $J_3$  as well. This can be naively connected to the emergence of peaks in the three-spin Ising model environment. The generation and fall of concurrence is a combined effect of various decoherence channels that we will explore in Appendix  \ref{appA}. The  effect of $J_3$ clearly causes an interesting time profile of $|d^F_{\alpha \beta,\lambda \gamma}(t)|^2$ as compared to TFIM environment as far as their relative decays are concerned \cite{PhysRevA.94.022316}. } One can get the underlying Schmidt rank of the time-evolved state $\rho^F_S(t)$ greater than $2$ due to the finite weight of the off-diagonal terms in early time. However, Schmidt rank approaches unity as time progresses, resulting in a factorized state of the central spins at later time.

Now coming to Figs. \ref{fig:Dynamic_collage} (a,b) for $d=0$ and $d=5$, respectively, with  
weak three-spin coupling (\( J_3 = 0.1 \)),
we find that bi-partite entanglement between the two central spins remains non-zero for longer times if the coupled environmental spins are spatially separated. This behavior can be attributed to the fact that $d^F_{\downarrow \uparrow, \uparrow \downarrow}$ stays at unity for all time when $d=0$. Since  $d^F_{\downarrow \uparrow, \uparrow \downarrow}$ varies with time for $d\ne 0$, the secondary fall of the concurrence is faster for the spatially separated case. This is generically observed in  Figs. \ref{fig:Dynamic_collage} (c,d) with $J_3=0.6$, Figs. \ref{fig:Dynamic_collage} (e,f) with $J_3=1.5$, and Figs. \ref{fig:Dynamic_collage} (g,h) with $J_3=2.0$.  Importantly, in terms of the generation of bi-partite entanglement, the intra-phase quench is less effective as compared to the inter-phase quench, see Figs. \ref{fig:Dynamic_collage} (e,f,g,h). This also naively explains the slower growth of concurrence for the intra-phase quench, leading to a large value of $t_1$ as compared to the inter-phase quench.

Another interesting observation is that when an inter-phase sudden quench across a quantum critical point with $J_3-1<h_I<J_3+1$ and $h_3=2.2>J_3+1$, the two-stage fall of concurrence is observed. If $h_I$ is chosen from the vicinity of a critical point, i.e, $h_I=0.6 \approx J_3-1$ with $J_3=1.5$, a two-stage fall converts into a single-stage fall, see Figs. \ref{fig:Dynamic_collage} (e,f). A similar phenomenon is observed in Figs. \ref{fig:Dynamic_collage} (g,h) for $h_I=1$ while $J_3=2$. 
In the case of intra-phase sudden quench without crossing a quantum critical point, i.e, $h_I=1$, [$h_I=1.6$] for $J_3=1.5$ [$J_3=2.0$] keeping  $h_F=2.2$ as shown in Figs. \ref{fig:Dynamic_collage} (e,f) [Figs. \ref{fig:Dynamic_collage} (g,h)], the first fall of the concurrence is substantially slow, giving rise to a prolonged behavior of concurrence with time. This is quantitatively confirmed by our scaling analysis, which finds significantly suppressed exponents for intra-phase quenches, with $\beta_1$ in the range $0<\beta_1<0.3$  and $\beta_2$ in the range $0.5 <\beta_2<1.2$. Here, the secondary peak time $t_2$ is much larger compared to other quench schemes. This is followed by the secondary fall, causing the concurrence to remain finite for a long time. Therefore, inter(intra)-phase quench can induce stronger (long-lived) bi-partite entanglement between the central spins as compared to intra(inter)-phase quench schemes. The magnetic nature of the initial or final phases, associated with the quench, does not have many distinct effects on the dynamics of the concurrence.

The results above are markedly distinct compared to those of the TFIM environmental spin chain when $J_3$ increases. For example, the long-lived nature of concurrence, observed for intra-phase quench, is commonly observed for both TFIM and three-spin Ising environment when $J_3 \to 0$. The long-lived profile of concurrence varies significantly with $J_3$ where the first fall between primary and secondary peaks is indeed slow as compared to the TFIM counterpart. Also, the single-stage fall of concurrence for the critical quench in the present case is not noticed in the TFIM environmental chain. Therefore, three-spin interaction $J_3>0.5$ can substantially impact the dynamics of the decoherence channels, which would result in distinct properties of entanglement growth and subsequent decay. Note that the three-spin dominated region appears in the phase diagram when $J_3>0$.

\begin{figure}[ht]
\centering
          \includegraphics[width=0.42\textwidth]{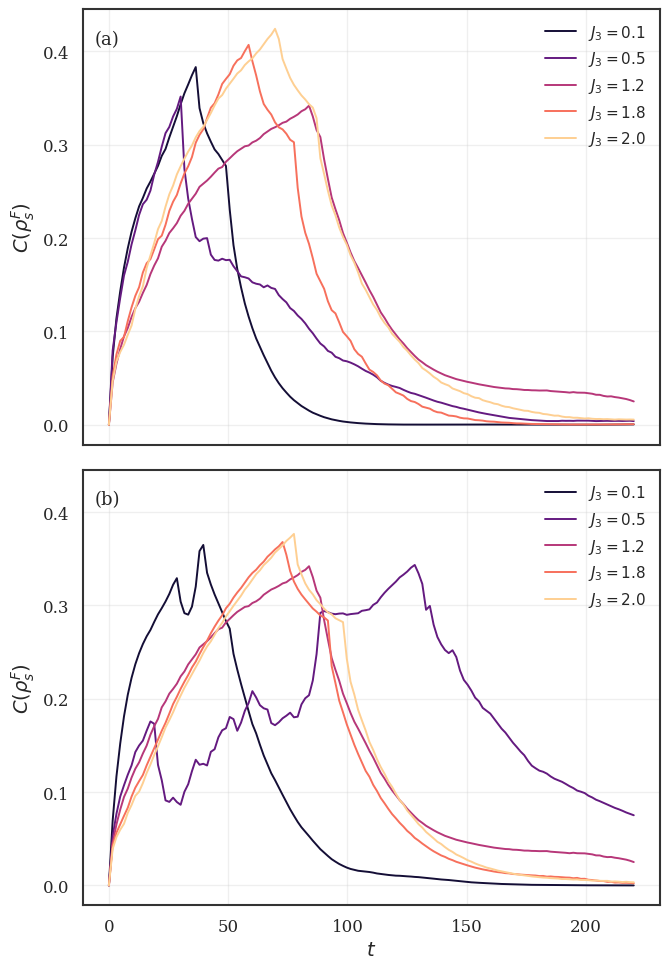}
\caption{ We show the dynamics of concurrence $C(\rho_S^F)$  with $h_I=-1.5$, $1.5$ and $h_F=1.5$,$-1.5$ in (a) and (b), respectively. We consider $J_3=0.5$ (black), $1.2$ (purple), $1.8$ (red), $2.0$ (yellow) with $d=0$. We consider $N=20$ for exact diagonalization calculation.}
\label{fig:reversalhf}
\end{figure}



\subsubsection{Concurrence under the sign reversal of $h_F$}
\label{secb1}

In order to investigate the effect of $J_3$ more extensively, we analyze the reversal of the transverse field $h$ from $h \to -h$, see Fig. \ref{fig:reversalhf}.
One can find that the peak value of the concurrence 
as well as the primary peak time instant $t_1$ do not have any systematic correlation with $J_3$ for $h_F=1.5$, while both of the above quantities  
increase as $J_3$ increases for $h_F=-1.5$, see Figs. \ref{fig:reversalhf} (a,b). This clearly suggests that the sign of $h_F$ makes noticeable differences in the behavior of concurrence when $J_3$ is large.  We also note that concurrence is found to acquire a higher value for negative $h_F$ as shown in Fig. \ref{fig:phase-diagramconc1}. This may cause bi-partite entanglement between the central spins to achieve a relatively higher value for negative $h_F$ as compared to positive $h_F$. 
Importantly, in the TFIM environment, such a reversal of the transverse field would not affect the dynamics of entanglement between the central spins, such that it shows qualitatively distinct features.   This can be understood from the invariance of the TFIM Hamiltonian under $h\to -h$ and $\sigma_z \to -\sigma_z$ leading to no change in the equilibrium phase diagram. However, this is not the case for the three-spin model as the Hamiltonian does not remain invariant under $h\to -h$ and $\sigma_z \to -\sigma_z$ due to the finite value of $J_3$. The dynamics of concurrence are markedly different under the reversal of the magnetic field when $J_3>0.5$.  


\subsubsection{Dynamical Landscape of Central Spin Entanglement}
\label{sec3.5}

\begin{figure}[ht]
\centering
          \includegraphics[width=0.45\textwidth]{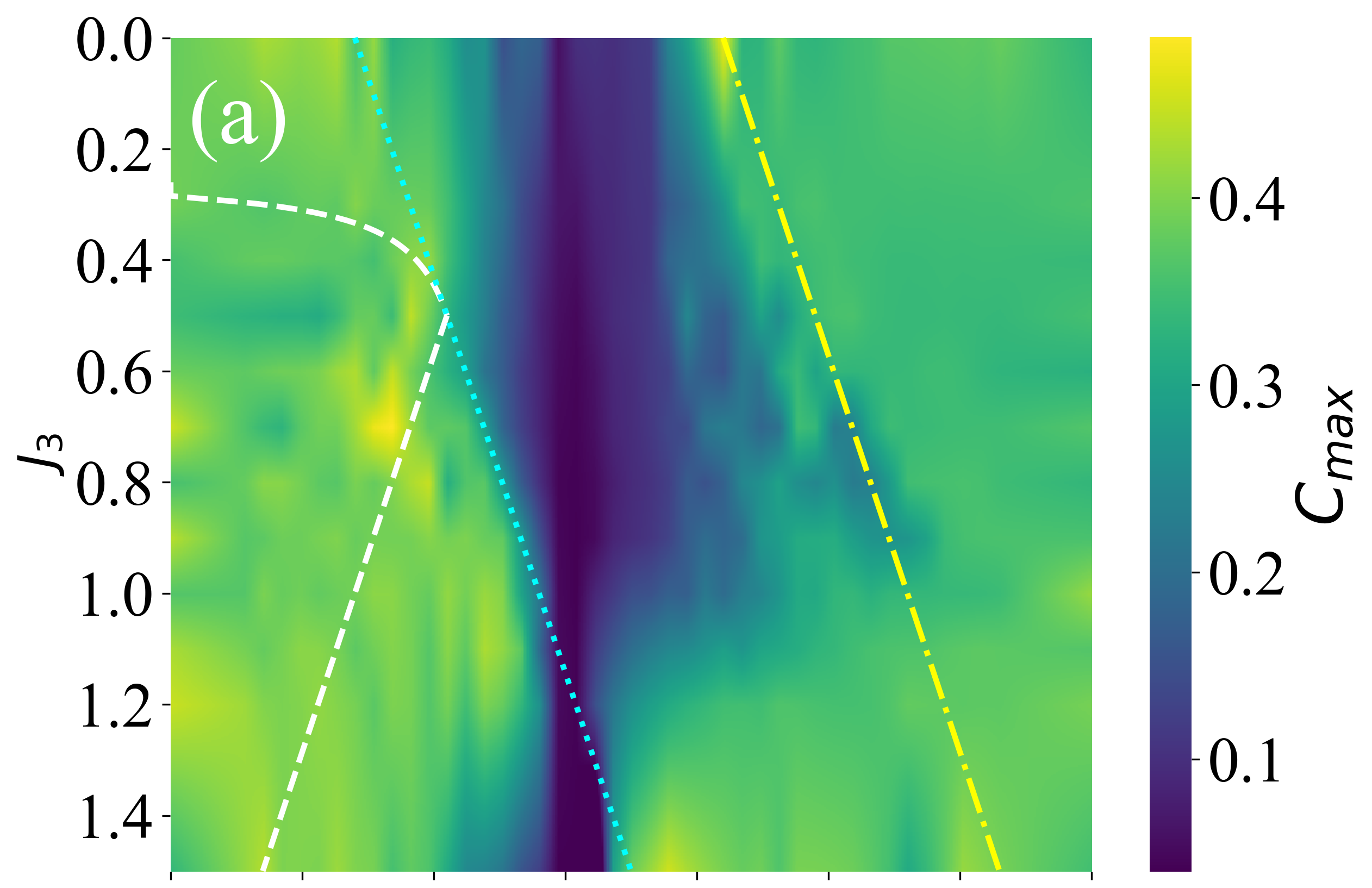}
          \includegraphics[width=0.45\textwidth]{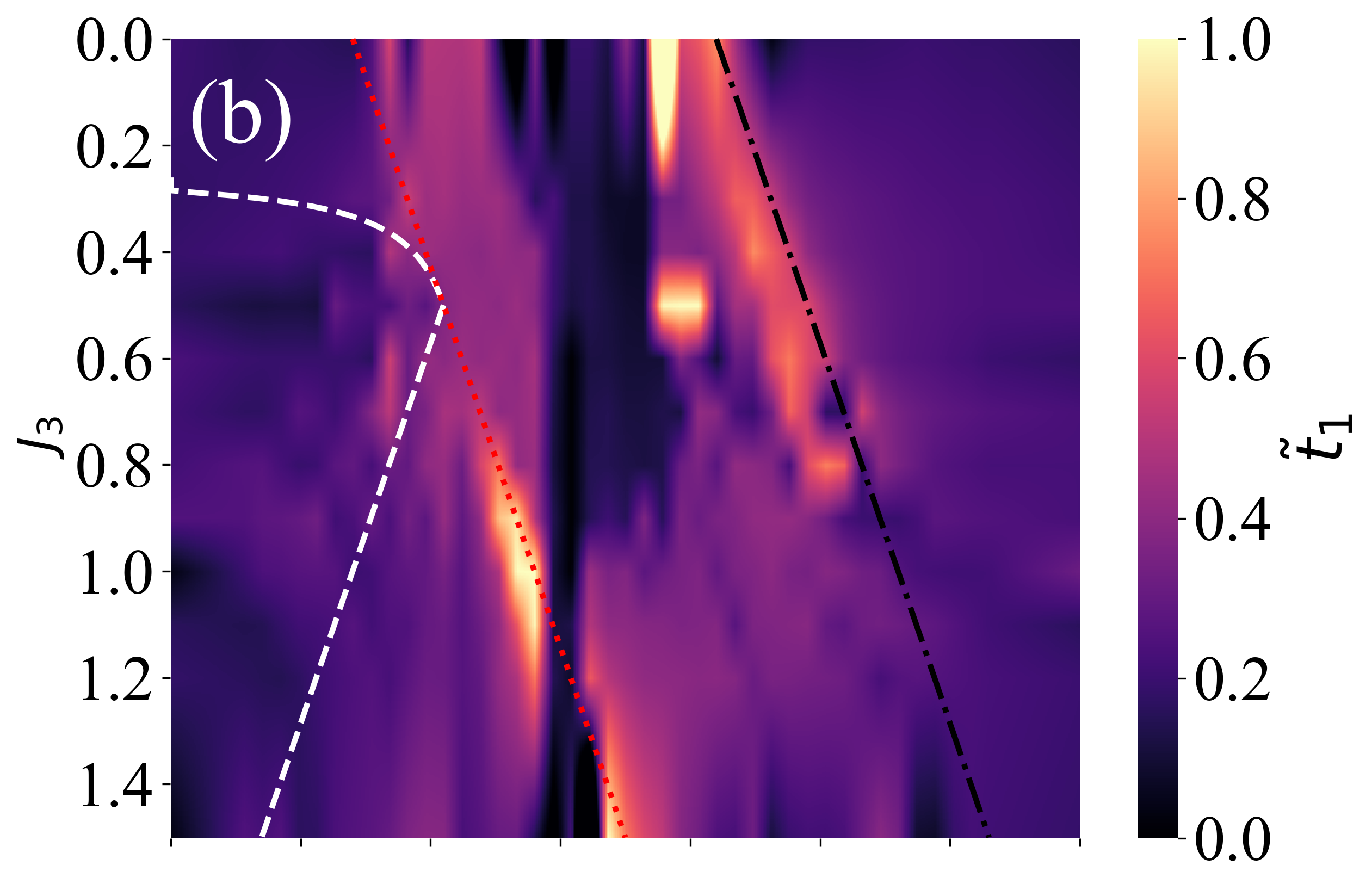}
          \includegraphics[width=0.45\textwidth]{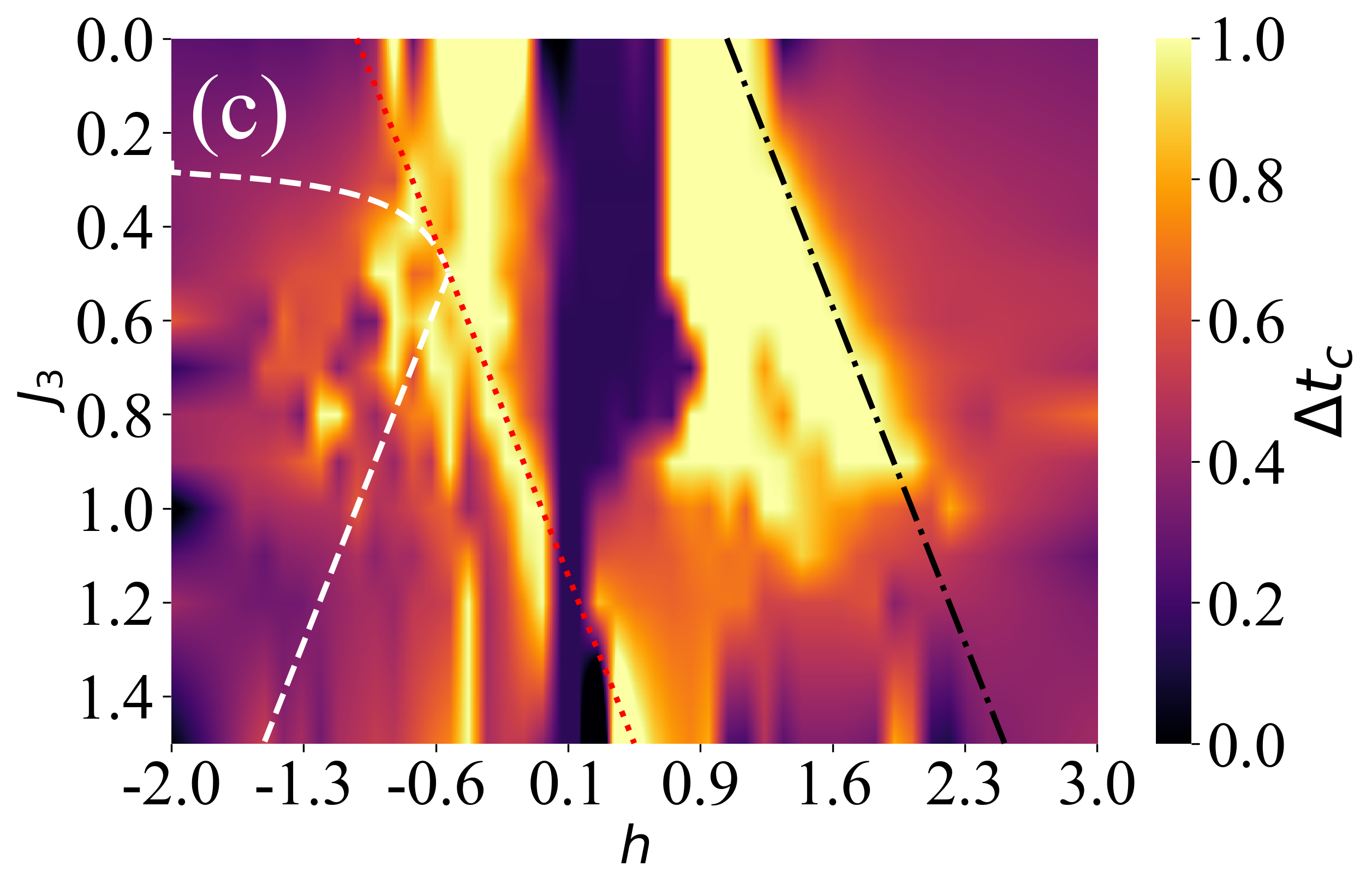}
\caption{We show the density plots for the peak value of concurrence $C_{\rm max}$ in (a),  normalized time $\tilde{t}_1$ at which concurrence reaches its peak in (b), and the normalized lifetime $\Delta t_c$ of concurrence i.e., the duration for which it remains non-zero in (c) over $h_F$-$J_3$ plane with $h_I=0.2$, $d=0$. Dashed-dotted line corresponds to line $h=J_3+1$, dotted line corresponds to $h=J_3-1$ and the dashed line corresponds to $h=-J_3$ for $J_3>0.5$ and  $h=\frac{J_3}{1 - 4J_3}$ for $0.25<J_3\le0.5$.  We have considered $N=16$ spins for these calculations. } 
 
\label{fig:concurrence-collage}
\end{figure}

To obtain a global perspective on the entanglement behavior between two central spins, we construct dynamical phase diagrams using concurrence $C(\rho_S^F)$ that data sampled across a broad range of final value of transverse field values \( h_F \) and three-spin interaction strengths \( J_3 \) while $h_I=0.2$ is kept fixed, see Fig. \ref{fig:concurrence-collage}. These diagrams provide a statistical and time-resolved overview of how central spin entanglement evolves throughout the parameter space, highlighting regions of enhanced quantum correlations and revealing the interplay between three-spin interaction and transverse field.

We show the maximum value of the concurrence $C_{\rm max}$ in Fig.~\ref{fig:concurrence-collage}(a) observed during time evolution for each pair \((h_F, J_3)\). The heatmap reveals regions of enhanced entanglement generation, with high concurrence values clustering around the
critical line $h_F=J_3-1$. In particular, the asymmetry under \( h_F \leftrightarrow -h_F \) is evident, especially at intermediate and high values of \( J_3 \). Interestingly, a strong entanglement ridge, captured by a bright yellow region, is visible near the multicritical critical point $h_F=J_3-1$ and $h=-J_3$, indicating critical enhancement in that region. A dark blue patch is observed in the region $h_F \in [0.2, 0.4]$, indicating that the generation of concurrence is suppressed for intra-phase quench. 
We observe a greenish-blue patch along the $h=J_3+1$ phase boundary. In this region, the concurrence peak is comparatively suppressed, indicating weaker entanglement enhancement relative to the other critical line $h=J_3-1$ which is in the vicinity of the three-spin interaction dominated region. This heatmap explains the effect of $J_3$ and the role of critical points to generate a substantial bi-partite entanglement between the central spins.

We now illustrate the heatmap of normalized time $\tilde{t}_1$ at which the concurrence reaches its primary maximum in Fig.~\ref{fig:concurrence-collage}(b) over the $h_F$-$J_3$ plane. What we have observed from Fig. \ref{fig:Dynamic_collage} is that peak time $\tilde{t}_1$ increases for intra-phase quench, leading to an extended tail of concurrence. This is clearly observed between the critical lines $h_F=J_3-1$ and $h_F=J_3+1$ when the brighter regions appear. The three-spin interaction marks its impression between $h_F>-J_3$ and $h_F<J_3-1$ for $J_3>0.5$, which is relatively brighter than the region $h_F>J_3+1$ and $h_F<J_3-1$ for $J_3<0.5$. 
Regions with early entanglement generation appear dark, matching expectations from dynamics when $h_F \approx h_I$. 
We also observe a clear delay in the primary peak of concurrence with time around the vicinity of the critical line $h_F=J_3\pm 1$. The most delayed entanglement generation takes place around  $h_F=J_3+1$ with $J_3 \approx 0.5$ and around $h_F=J_3-1$ with $J_3 \approx 1$. These bright regions are associated with the intra-phase quench only and hence validate the finding of larger $\tilde{t}_1$ for intra-phase quench.

We now probe the normalized time window $\Delta t_c$ within which two central spins remain entangled, namely, the  
persistence of entanglement in Fig.~\ref{fig:concurrence-collage}(c). This heatmap closely follows the Fig.  \ref{fig:concurrence-collage}(b) as the delay in the primary peak eventually results in a long tail of entanglement. 
Extended durations of concurrence are observed in wide parameter ranges, particularly when $h_F \in [0.5, 1.8]$ and for moderate $J_3<1$ such that $h_F<J_3+1$. Around the critical line $h_F=J_3-1$, and multicritical point $h_F=-J_3=J_3-1$, one can find bright spots indicating the long-lived behavior of entanglement.  The dark vertical stripe near $h_F \approx h_I$ corresponds to the absence of entanglement generation during the time evolution.  The reddish-yellow region is found to be 
inside the phase boundaries $h_F=J_3 \pm 1$ clearly signifying 
the effect of $J_3$  to sustain the entanglement for a longer time. The blackish-red zones beyond the phase boundaries indicate the short-lived nature of the entanglement.

Now, combining the three heatmaps, we can comment that 
the strength of entanglement is maximized for a critical quench around the multicritical point when $J_3>0.5$ while the growth (sustainability) of entanglement is slower (wider)  for the intra-phase quench, and both these characteristics are sensitive to the critical line $h_F =J_3\pm 1$ for non-zero values of $J_3$.  Therefore, quenching close to the vicinity of the multi-critical point is the most optimal for generating high entanglement with long-lived character. On the other hand, entanglement generation is significantly suppressed if $h_F$ is chosen in the vicinity of $h_I$.

\section{Conclusions}
\label{sec4}

We consider a three-spin interacting Ising model where we study the bi-partite entanglement between adjacent spins, quantified through concurrence, which exhibits peak-valley structure signifying the critical and off-critical nature of the model. Interestingly, the three-spin dominated paramagnetic region exhibits vanishing concurrence that can be attributed to the tri-partite nature of the underlying phase. We next consider a generalized central spin model where two central spins are locally coupled to an environmental three-spin interacting Ising model, to study the generation of bi-partite entanglement.  We examine the equilibrium (non-equilibrium) case where the transverse field is kept fixed (suddenly changed) once the coupling between the central spins and
the environmental spin chain is established. 
The initial two-qubit pure state of the central spins becomes a mixed entangled Werner state with time, where the decoherence channels play a significant role in determining the profile of the concurrence of the central spins. The weak (strong) nature of decoherence channels causes a minimal (substantial) amount of bi-partite entanglement to be induced between the spins for equilibrium (non-equilibrium) cases. The equilibrium profile of central spin concurrence is governed by quasi-particle dynamics, and their velocities clearly dictate the temporal location of the dip in the profile.

In the case of non-equilibrium dynamics, we find a distinct decay profile of concurrence between intra- and inter-phase quenching while the growth profile is found to be qualitatively unaltered.  
The primary peak of concurrence is associated with the complete suppression of a particular decoherence channel. The two central spins become maximally entangled for a quench in the vicinity of the critical line, especially near the multicritical point, which clearly signifies the non-trivial role of three-spin interaction. The two-stage fall is not impacted by the critical lines. The intra-phase quench results in slow generation of entanglement, followed by a long tail enforcing a long-lived entanglement. This behavior is also quantitatively tuned by varying the strength of the three-spin interaction. Remarkably, the entanglement generation strongly depends on the sign of the transverse field in the presence of the three-spin interaction. Therefore, the three-spin term of the environment plays a crucial role in generating, enhancing and sustaining the bi-partite entanglement between the central spins.
Given the experimental advancement in optical lattices and trapped ion setups for realizing different spin models \cite{Bloch2012,Simon2011,Kim2010}, and experimental  detection of entanglement in a central spin model using nuclear spin-bath \cite{Maze2008,science1131871,PhysRevLett.103.040502}, we believe that our work is experimentally viable. In the future, one can study the effect of frustration present in the environmental spin chain on the dynamics of entanglement.

\appendix

\section{Analysis Through Decoherence Channels}
\label{appA}


\begin{figure}[ht]
\centering
          \includegraphics[width=0.43\textwidth,height=8.5cm]{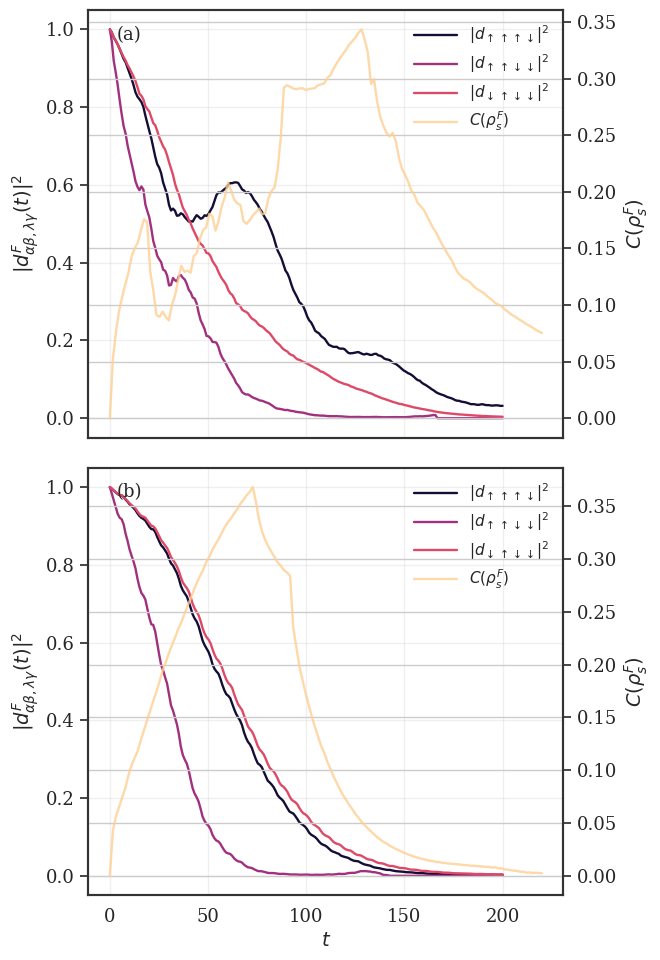}
\caption{We show the evolution of decoherence channels $|d^F_{\alpha \beta, \lambda \gamma}(t)|^2$ (left axis) and concurrence $C(\rho_S^F)$ (right axis)  with $h_I=-1.5$, $h_F=1.5$ , $J_3=0.5$ and $d=0$ in (a) and $h_I=1.5$, $h_F=-1.5$, , $J_3=1.8$ and $d=0$ in (b). We consider  $N=20$ spins for exact diagonalization. }
\label{fig:decoherence}
\end{figure}


\textcolor{black}{We here simultaneously examine the time evolution of $|d_{\alpha\beta,\lambda\gamma}|^2$ and concurrence $C(\rho_s^F)$. 
For $h_I = 0.2$ and $h_F = -1.5$ with $J_3 = 0.5$ and $1.8$ [Fig.~\ref{fig:decoherence}(a,b)], we observe that the temporal features of the decoherence channels and the concurrence exhibit a possible correlation. In particular, the concurrence maxima occur in time windows where certain dominant off-diagonal channels, notably $|d_{\uparrow\uparrow,\downarrow\downarrow}|^2$, become strongly suppressed. We emphasize, however, that this correspondence is purely empirical and does not imply a direct or causal relation. Since the concurrence depends nonlinearly on all elements of the reduced density matrix, no single channel determines its behavior. Rather, the concurrence profile reflects the collective evolution and interference of all decoherence channels. As the channels progressively decay at long times, the reduced state approaches the maximally mixed state $\rho^F_s (t\to \infty)=I/4$, resulting in the eventual disappearance of entanglement. We further observe numerically that increasing $J_3$ leads to smoother decay profiles of the individual channels. In such cases, the concurrence growth becomes more gradual and can exhibit a more monotonic temporal increase, although this behavior remains a global consequence of the full density-matrix dynamics rather than of any specific channel.}

\subsection*{Data Availability}
The data supporting the findings of this article are available upon request to the authors.

\acknowledgments
T.N. acknowledges the NFSG ``NFSG/HYD/2023/H0911'' from BITS Pilani. T.N. thanks late Amit Dutta for introducing him to the generalized central spin model.  We thank  SARANGA: High-Performance Computing Facility provided by BITS Pilani-Hyderabad campus. We thank  Sarmistha Banik and the Physics Department for providing additional computational support.

%
\end{document}